\documentclass[%
 aip,
 amsmath,amssymb,
 reprint,%
]{revtex4-1}
\usepackage{siunitx}
\usepackage{graphicx}
\usepackage{dcolumn}
\usepackage{bm}
\usepackage[hidelinks]{hyperref}
\hypersetup{
pdfcreator={},
pdfproducer={}
}
\usepackage[utf8]{inputenc}
\usepackage[T1]{fontenc}
\usepackage{mathptmx}

\begin{document}

\preprint{AIP/123-QED}

\title[Photon flux determination of a liquid-metal jet x-ray source by means of photon scattering]{Photon flux determination of a liquid-metal jet x-ray source by means of photon scattering}

\author{Malte Wansleben}
 \email{malte.wansleben@ptb.de}
\author{Claudia Zech}
\author{Cornelia Streeck}
\author{Jan Weser}
\affiliation{ 
Physikalisch-Technische Bundesanstalt (PTB), Abbestraße 2-12, 10587 Berlin, Germany
}%

\author{Christoph Genzel}
\affiliation{%
Helmholtz-Zentrum Berlin für Materialien und Energie GmbH (HZB), Hahn-Meitner-Platz 1, 14109 Berlin, Germany
}%

\author{Burkhard Beckhoff}
\affiliation{ 
Physikalisch-Technische Bundesanstalt (PTB), Abbestraße 2-12, 10587 Berlin, Germany
}%
\author{Roland Mainz}
    \email{roland.mainz@helmholtz-berlin.de}
\affiliation{%
Helmholtz-Zentrum Berlin für Materialien und Energie GmbH (HZB), Hahn-Meitner-Platz 1, 14109 Berlin, Germany
}%

\date{\today}

\begin{abstract}
Liquid-metal jet X-ray sources promise to deliver high photon fluxes, which are unprecedented for laboratory based X-ray sources, because the regenerating liquid-metal anode is less sensitive to damage caused by an increased electron beam power density. For some quantitative X-ray analysis techniques, knowledge of the absolute photon flux is needed. However, a precise experimental determination of the photon flux of high-performance X-ray sources is challenging, because a direct measurement results in significant dead time losses in the detector or could even harm the detector. Indirect determinations rely on data base values of attenuation or scattering cross sections leading to large uncertainties. In this study we present an experimental determination of the photon flux of a liquid-metal jet X-ray source by means of elastic and inelastic photon scattering. Our approach allows for referencing the unpolarized output radiation of the liquid-metal jet X-ray source to polarized synchrotron radiation in a simple setup. Absolute photon fluxes per solid angle are presented with a detailed uncertainty budget for the characteristic emission lines of Ga K\(\alpha\) and In K\(\alpha\) for two different acceleration voltages of the X-ray source. For a nominal setting of 200 W (70 kV, 2.857 mA) the determined values for Ga K\(\alpha\) and In K\(\alpha\) are 6.0(5)\(\times 10^{12}\) s\(^{-1}\)sr\(^{-1}\) and 3.8(4)\(\times 10^{11}\) s\(^{-1}\)sr\(^{-1}\), respectively. 
\end{abstract}

\maketitle

%

Electron-impact X-ray sources are compact solutions for laboratory based research and applications. As the brilliance scales with the electron-beam power density at the anode, the physical requirements of the anode material become more challenging since thermal stress and damage at the anode surface degrade the output power. The implementation of a rotating anode overcomes this quandary partly, however, intrinsic thermal limitations are still present. A more recent concept is the use of a liquid-metal jet anode material\cite{Hemberg2003}. The fluid anode jet is capable of taking a higher thermal load. Additionally, the increased velocities of the jet (up to 100 m/s) compared to a rotating anode provide a faster regeneration and perpetual smooth anode surface\cite{Hemberg2003, Hallstedt2018}. The X-ray flux produced by the liquid-metal jet X-ray source (LMJ) is expected to come close to bending-magnet synchrotron sources\cite{Otendal2008}, 
enabling numerous applications such as high-resolution imaging\cite{Larsson2011}, X-ray microscopy\cite{Thuring2013}, X-ray phase-contrast tomography \cite{Larsson2016}, confocal micro X-ray fluorescence spectroscopy\cite{Bauer2018}, or real-time X-ray diffraction in the laboratory. 

The knowledge of the incident X-ray flux is needed for quantitative analyses\cite{Beckhoff2008a}, however, a direct experimental determination of high X-ray fluxes is troublesome because conventional X-ray spectrometers are not capable of dealing with count rates of the above-mentioned magnitude resulting in significant dead time losses or even radiation damage. Hence, there are various indirect experimental or semi-empirical approaches to estimate the output spectrum of X-ray tubes, such as Compton scattering \cite{Yaffe1976, Matscheko1987, Maeda2005}, induced fluorescence emission in targets \cite{Moran1985}, combined attenuation and scattering of the primary beam \cite{Mainardi1989}, combined scattering experiments and Monte Carlo simulation \cite{Padilla2005}, or attenuation of the scattered beam \cite{Mainardi2008}. All of these approaches rely on data base values describing the respective physical processes involved in the photon matter interaction, such as the attenuation coefficient or scattering cross sections. Hence, experimental results strongly depend on these parameters and their uncertainties. Although the PTB has access to calibrated silicon drift detectors (SDD)\cite{Beckhoff2000}, a direct determination with the detector directed towards the LMJ is not feasible. The main reason is the expected high photon flux which would require appropriate absorbers to prevent the detector from exposure to high photon fluxes. Due to the strong energy-dependence of the attenuation coefficients, it is not possible to sufficiently attenuate the high-energy range of the radiation without completely suppressing the low-energy range.

In this work we present an indirect approach that allows to determine the absolute photon flux with low uncertainty based on referencing detected elastically and inelastically scattered photons measured at the LMJ to scattered photons measured at a synchrotron radiation beamline. This method is independent of tabulated scattering cross sections. Only the attenuation coefficient for air is needed. We apply this method to determine the fluxes of the characteristic K\(\alpha\) emission lines of gallium and indium which are the main constituents of the LMJ anode. 

\begin{figure}
\centering
\includegraphics[width=1.0\linewidth]{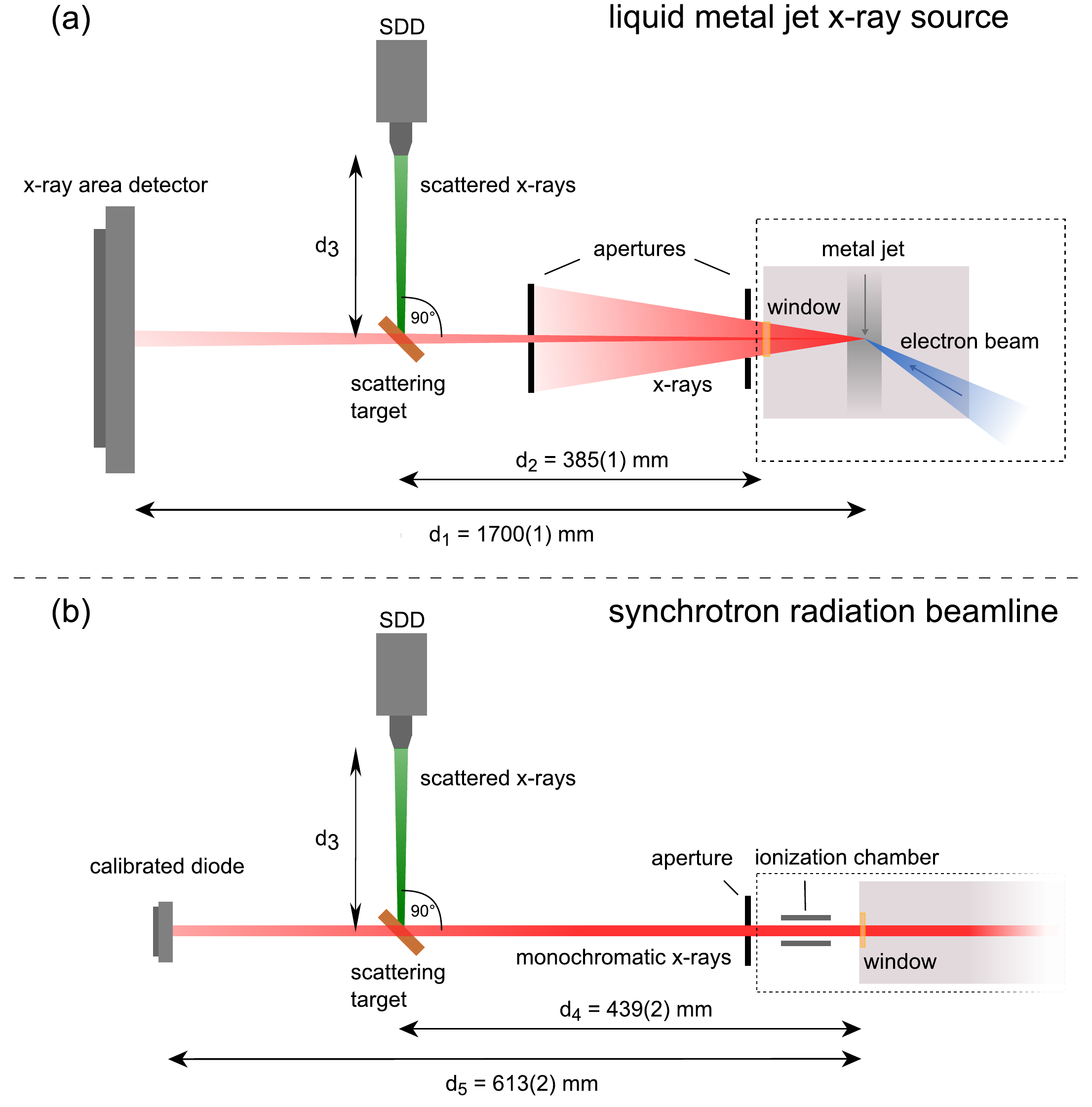}
\caption{Schematic side view of the experimental setup using (a) the metal-jet X-ray source and (b) the synchrotron radiation beamline (BAMline) at BESSY II. The scattering angle \(\theta = 90^{\circ}\) is kept constant in both experiments. The azimuthal angle \(\phi\) is changed from \(0^{\circ}\) to \(90^{\circ}\).}
\label{fig:LMJ_SDD_Scanner}
\end{figure}

\begin{figure}
\centering
\includegraphics[width=1.0\linewidth]{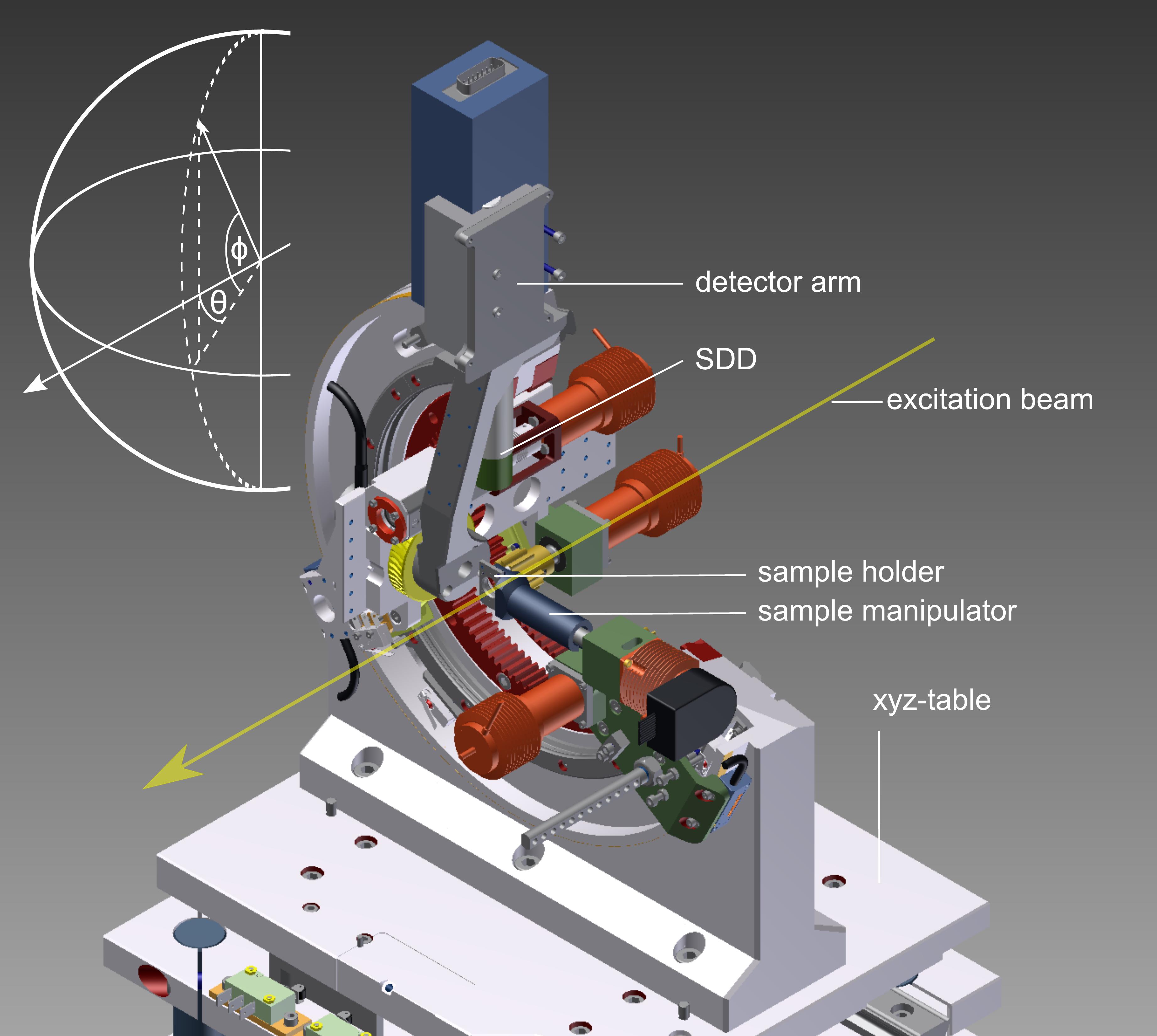}
\caption{Schematic of the SDD scanner end-station. A declaration of scattering angle \(\theta\) and azimuthal angle \(\phi\) with respect to the incident excitation beam is given in the upper left corner. }
\label{fig:SDD_scanner}
\end{figure}

A schematic view of the two experimental setups is shown in Fig. \ref{fig:LMJ_SDD_Scanner}. In both setups the same experimental end-station, the so-called SDD scanner, is used to relate the detected count rate of the scattered characteristic emission lines of the LMJ to the scattered monochromatic synchrotron radiation. The SDD scanner allows for the recording of an X-ray spectrum in almost any combination of incident and exit angle. A schematic of the SDD scanner is depicted in Fig. \ref{fig:SDD_scanner}. The SDD scanner comprises a sample manipulator and a detector arm and allows for an independent alignment of the surface normal of the sample and the detector front-end with any unit vector of the spherical coordinate system, respectively. 
While the sample rests in the center of the common spherical coordinate system, the SDD on the detector arm is allowed to move on a spherical surface around this center. The rotational degrees of freedom are fully motorized. The angular resolution is approximately 3\(^{\circ}\). The SDD-scanner was designed for the experimental determination of scattering cross section and spectroscopic analysis of light elements in the hard X-ray regime.

The anode material (ExAlloy I1) of the used LMJ from Excillium (model D2) contains nominal weight fractions of 68\% gallium (Ga), 22\% indium (In) and 10\% tin (Sn). The dimension of the electron beam was nominally set to 50 \(\mu\)m high and 80 \(\mu\)m wide resulting in a projected size of 50 \(\mu\)m $\times$ 20 \(\mu\)m according to the manufacturer. We note that the projected spot size can be further reduced to 20 \(\mu\)m $\times$ 20 \(\mu\)m with the same power settings. The electron impact ionizes the atoms of the jet which results in the isotropic emission of characteristic X-ray fluorescence. An exit window defines the emission direction and separates the vacuum chamber from the ambient condition in the laboratory. Two apertures behind the exit window collimate the emitted X-rays towards the experiment. In order to determine the solid angle $\Omega_{L}$ of emission defined by the two apertures, an image of the circular excitation beam profile is recorded by means of an X-ray area detector at distance \(d_1\) from the source point (Fig. \ref{fig:LMJ_SDD_Scanner}(a)). The beam radius \(r\) at the detector position is determined by considering the pixel size of the detector leading to a solid angle of $\Omega_{L} = 2.5(1) \times 10^{-5} \text{sr}$. Thus, at the position of the scattering target, the diameter of the beam corresponds to 0.74(2) mm.

The scattering experiment with the SDD scanner was repeated with monochromatic synchrotron radiation provided by the 7-T wavelength shifter (WLS) beamline (BAMline)\cite{Gorner2001, Riesemeier2005} at the electron storage ring BESSY II. The BAMline is equipped with a double-crystal monochromator (DCM) and a double multilayer monochromator (DMM) which was operated in series. While the DCM provides high spectral resolving power, the DMM efficiently suppresses higher-order radiation below \(2\times 10^{-5}\). Energy calibration of the monochromators is realized with an energy scan across the K-absorption edge of a 12.5-\(\mu\)m-thick palladium foil for the excitation energy of 24.21 keV and 10-\(\mu\)m-thick copper foil for the excitation energy of 9.25 keV.
An experimental hutch enables experiments under ambient conditions. An ionization chamber is installed behind the exit window of the beamline for beam monitoring. A calibrated photodiode at distance \(d_5\) from the exit window is used to determine the incident radiant power which can then be converted into photon flux using the experimentally determined responsivity\cite{Gerlach2008} after the sample has been removed from the setup (Fig. \ref{fig:LMJ_SDD_Scanner} (b)). A set of apertures within the beamline was used to define a rectangular beam spot of 1\(\times\)1 mm\(^2\).

A 1.84-mm-thick soda-lime glass sample is used as scattering target. Depending on the incident energy and angle of detection, metal foils may exhibit interfering diffraction peaks as reported in Gerlach et al.\cite{Gerlach2009}. Glass is an amorphous material which prevents diffraction peaks. It mainly consists of silicone dioxide. Silicon and oxygen are light elements which do not exhibit any characteristic fluorescence lines in the examined spectral region of the characteristic emission lines of the LMJ. A disadvantage of light elements is the low scattering cross sections\cite{Chantler1995} requiring a sufficiently thick scattering target. 

Equal to any other electron impact based X-ray source, the emitted radiation of the LMJ is unpolarized with characteristic emission lines of the anode material accompanied by continuous Bremsstrahlung. This is in contrast to the BAMline providing highly monochromatized and polarized radiation within the storage ring plane of the synchrotron radiation facility BESSY II. Hence, the detected photons of the polarized excitation source has to be transformed such that it is comparable to the photons detected at the unpolarized LMJ. The definition of the differential scattering cross sections \( \mathrm{d} \sigma / \mathrm{d} \Omega \) for elastic and inelastic scattering of unpolarized (superscript \(U)\) and polarized radiation (superscirpt \(P\)) can be found elsewhere\cite{Cesareo1992, Schoonjans2011}. Conversion of the differential scattering cross section from polarized to unpolarized radiation demands an averaging over all possible azimuthal angles \(\phi\), which describes the angle out of the polarization plane of the synchrotron radiation (Fig. \ref{fig:SDD_scanner}). This can also be expressed in terms of the differential scattering cross section in the polarization plane (subscript \(\parallel\)) and out of the polarization plane (subscript \(\perp\))\cite{Cesareo1992, Roy1993}:
\begin{align}
\frac{\mathrm{d} \sigma^{U}}{\mathrm{d} \Omega} = \frac{1}{2} \left ( \frac{\mathrm{d} \sigma^{P}_\parallel }{\mathrm{d} \Omega} + \frac{\mathrm{d} \sigma^{P}_\perp}{\mathrm{d} \Omega} \right ), 
\label{eq:unpol_from_pol}
\end{align}
where subscript \(\parallel\) corresponds to \(\phi = 0^{\circ}\) and subscript \(\perp\) corresponds to \(\phi = 90^{\circ}\). 
For all measurements the scattering angle \(\theta = 90^{\circ}\) was fixed. Using a relatively thick scatterer, self-absorption is not neglectable. Thus, the surface normal of scattering target is set to  \(\theta = 135^{\circ}\) and \(\phi=45^{\circ}\), ensuring the same exit angle \(\alpha_{\text{out}} = 30^{\circ}\) for \(\phi=0^{\circ}\) and \(\phi=90^{\circ}\). The incident angle is \(\alpha_{\text{in}} =45^{\circ}\) in both cases. 

\begin{figure}
\centering
\includegraphics[width=1.0\linewidth]{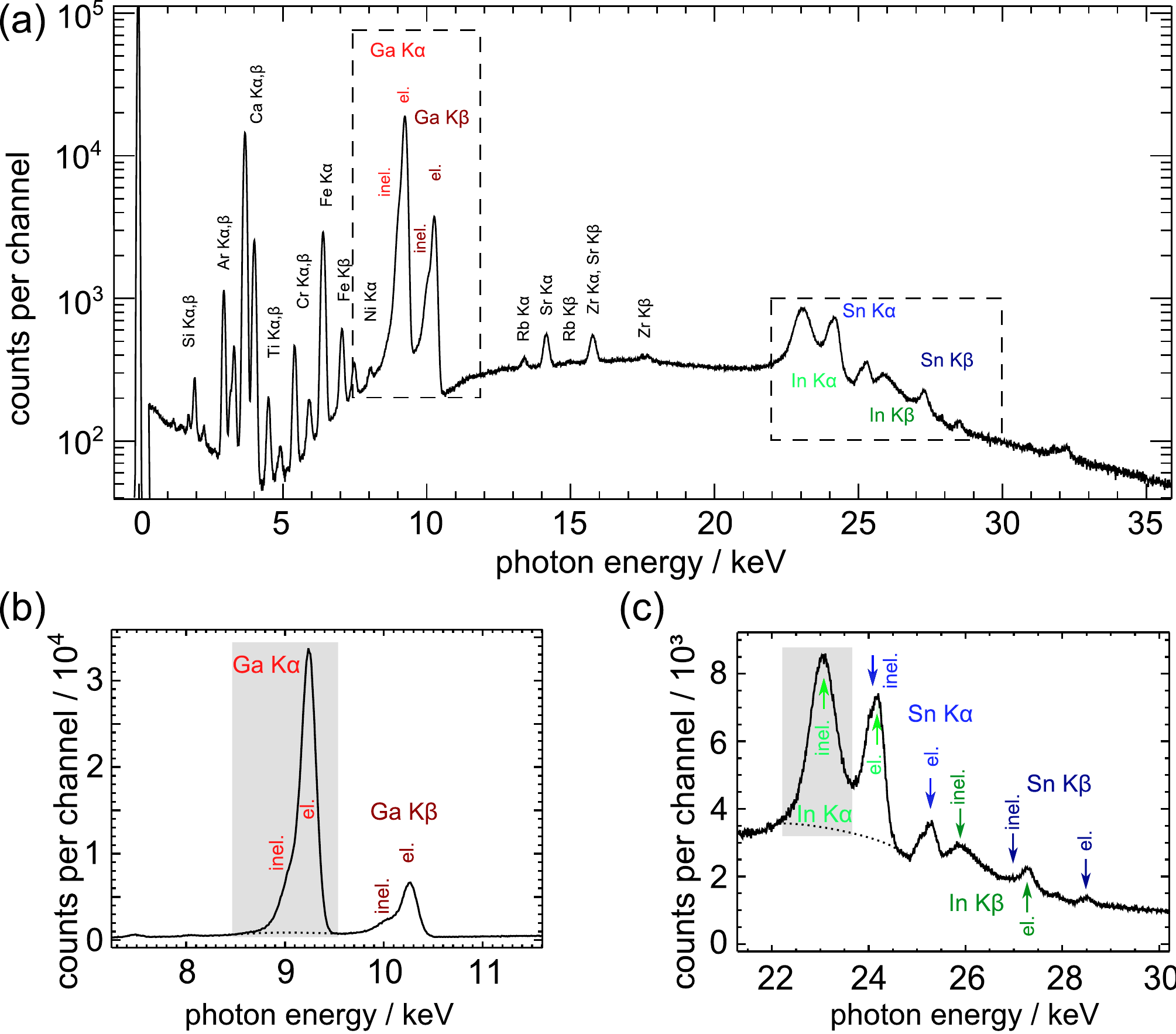}
\caption{SDD spectrum of the glass scattering target excited with the LMJ operated at 200 W (70 kV, 2.857 mA) and integration time of approx. 3h (11091 s). The SDD is positioned in the horizontal plane (\(\phi=0^{\circ}\)) with a scattering angle \(\theta =90^{\circ} \). Fluorescence lines of expected elements within the target, air and setup are also visible. The lines of interest are the elastic (el.) and inelastic (inel.) scattering peaks caused by the characteristic K\(\alpha\)- (red) and K\(\beta\)-lines (green) of the anode material of the LMJ, Ga (b) and In (c). }
\label{fig:LMJ_glass_spec}
\end{figure}

Figure \ref{fig:LMJ_glass_spec} (a) displays one of the two SDD spectra (\(\phi = 0^{\circ}\)) measured at the LMJ. Various fluorescence lines are observed including  argon (Ar) of the surrounding air and 3d metals (Ti, Cr, Fe, Ni) of the SDD scanner's steel construction caused by secondary excitation. The traces of rubidium (Rb), strontium (Sr) and zirconium (Zr) as well as large amounts of silicon (Si) and calcium (Ca) are found in the glass scattering target. The intensity of the Si peak is highly attenuated by the air distance between target and detector. The two highlighted areas mark the elastic and inelastic scattering signals of the characteristic emission lines of the liquid-metal anode. A more detailed view of these spectral regions is given in Fig.\ref{fig:LMJ_glass_spec} (b) and (c). 
The corresponding features of In and Sn are superimposed as indicated by the arrows in Fig.\ref{fig:LMJ_glass_spec} (c). The energy of the inelastically scattered photons is calculated according to Compton\cite{Compton1923}. Only the elastic count rate of Sn K\(\beta\) and the inelastic count rate of In K\(\alpha\) can be well separated. Hence, in order to calculate the photon flux of the characteristic In K\(\alpha\) line of the LMJ, only the inelastic In K\(\alpha\) is considered, whereas for Ga K\(\alpha\) line the sum of the inelastic and elastic peak is considered. The respective count rates are obtained by summing over the background subtracted region of interest as indicated by the grey shaded areas in Figs. \ref{fig:LMJ_glass_spec} (b) and (c). 

\begin{figure}
\centering
\includegraphics[width=1.0\linewidth]{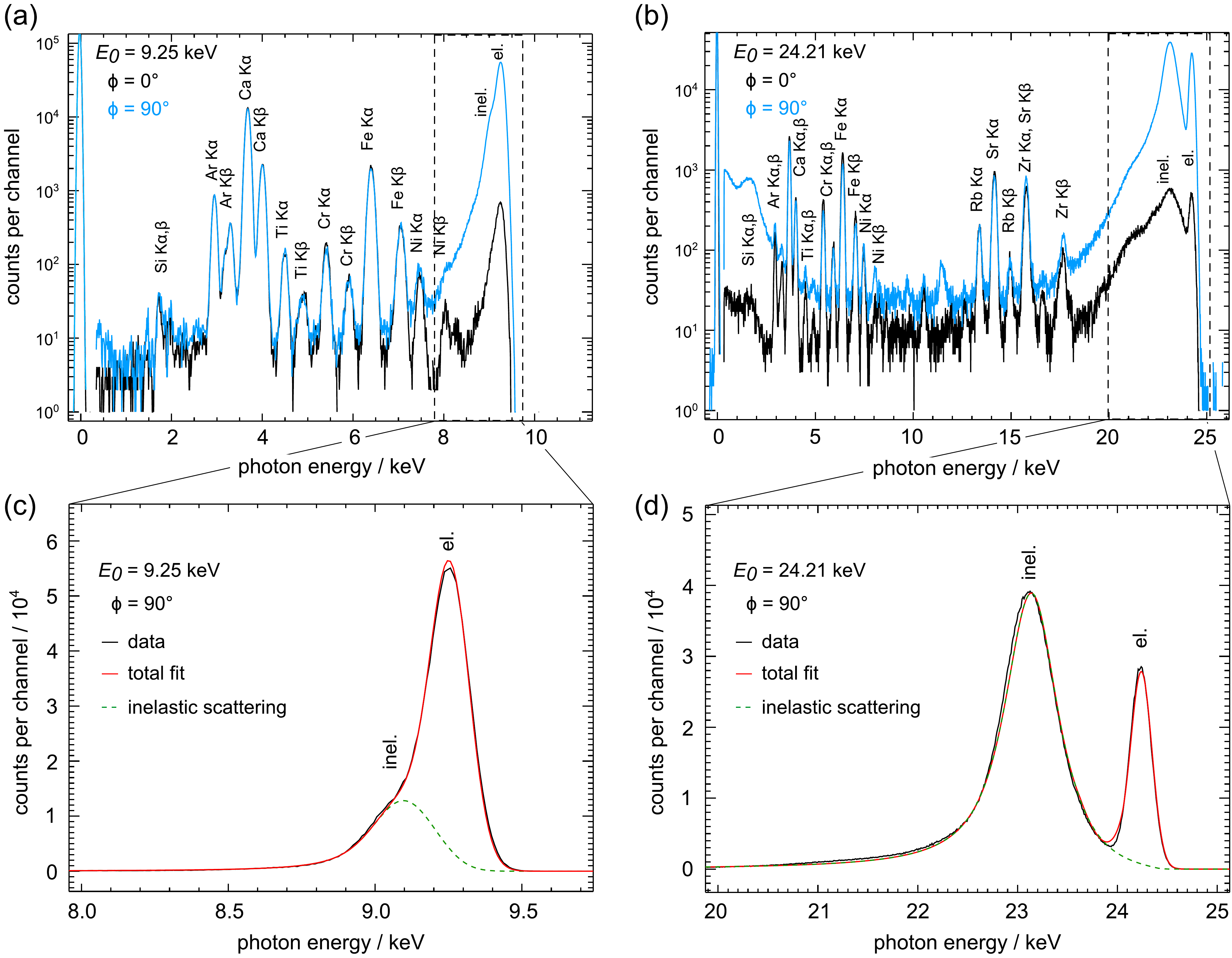}
\caption{SDD spectra of the glass scattering target excited with (a) 9.25 keV and (b) 24.21 keV at the BAMline. The spectra of the two measured azimuth angles \(\phi=0^{\circ}\) (black) and \(\phi=90^{\circ}\) (blue) are necessary to calculate the spectrum of an unpolarized excitation. The two dashed areas are dispalyed in figures (c) and (d) and show the deconvolution of the elastic and inelastic scattering peaks for the respective excitation energies. }
\label{fig:BAM_spec}
\end{figure}

Figures \ref{fig:BAM_spec} (a) and (b) show the SDD spectra recorded for excitation energies 9.25 keV and 24.21 keV for the two different azimuthal angles \(\phi\) at the BAMline. The count rates from both, elastic and inelastic scattering, are obtained from deconvolution of the spectra with a known spectrometer response\cite{Scholze2001}. The count rates associated with elastic and inelastic scattering increase with increasing azimuthal angle as expected due to the linearly polarized synchrotron radiation. The inelastic scattering cross section and the inelastic energy shift increase with increasing energy. Since elastic and inelastic peaks are hardly separated in the case of Ga, the count rate of both peaks are combined in a common region of interest. This is reasonable because the elastic and inelastic peaks are well separated from other fluorescence peaks in the spectra. 

For the unpolarized radiation of the LMJ the detected count rate $I_{\mathrm{L}}(E_{i})$
of the scattered photons with energy $E_{i}$ can be described by
the following equation derived from Henke et al.\cite{Henke1993}
\begin{eqnarray}
I_{\mathrm{L}}(E_{i})=I_{0,\mathrm{L}}^{U}(E_{0})A\left(d_{2}\right)M\frac{\mathrm{d}\sigma^{U}(E_{0},E_{i})}{\mathrm{d}\Omega}D.\label{eq:I_L}
\end{eqnarray}
$I_{0,\mathrm{L}}^{U}(E_{0})$ is the unpolarized photon flux of energy
$E_{0}$ leaving the LMJ in the solid angle defined by the aperture.
$A(d_{2})=\exp\left(-\mu_{\mathrm{air}}\rho_{\mathrm{air}}d_{2}\right)$
expresses the attenuation of the flux from the source window to the
scattering target (Fig. \ref{fig:LMJ_SDD_Scanner}(a)). $M\mathrm{d}\sigma/\mathrm{d}\Omega$
is the fraction of the incidence flux that is scattered in the direction
of the detector, where $M$ comprises the radiation path through the sample, its density, and self-attenuation effects. The factor $D$ summarizes all detector related quantities, such as detection efficiency, solid angle of detection, and air absorption between target and detector. %
The factors $M$ and $D$ are exactly the same for all measurements
at the LMJ and the synchrotron radiation source so they need not to be specified.
For elastic scattering $E_{i}=E_{0}$.

For the polarized radiation ($P$) of the synchrotron radiaition beamline we can express the count rates $I_{\mathrm{S}}^{||}$ (scattering in
the polarization plane) and $I_{\mathrm{S}}^{\perp}$ (scattering
out of the polarization plane) in a similar way:
\begin{eqnarray}
I_{\mathrm{S}}^{||,\perp}(E_{i})=I_{0,\mathrm{S}}^{P}(E_{0})A\left(d_{4}\right)M\frac{\mathrm{d}\sigma_{||,\perp}^{P}(E_{0},E_{i})}{\mathrm{d}\Omega}D.\label{eq:I_S}
\end{eqnarray}
$I_{0,\mathrm{S}}^{P}(E_{0})$ is the polarized flux at the window
of the beamline. For the calibration measurement by the photodiode
(Fig. \ref{fig:LMJ_SDD_Scanner}(b)), this flux is also attenuated by air absorption and hence
the derived photon flux is $I_{\mathrm{D,S}}^{P}=I_{0,\mathrm{S}}^{P}A(d_{5}).$ 

By inserting Eq. (\ref{eq:unpol_from_pol}) and Eqs. (\ref{eq:I_S})
into Eq. (\ref{eq:I_L}) we obtain the flux of the LMJ 
\begin{eqnarray}
I_{0,\mathrm{L}}^{U}(E_{0})=\frac{2I_{\mathrm{L}}(E_{i})I_{\mathrm{D,S}}^{P}(E_{0})}{I_{\mathrm{S}}^{||}(E_{i})+I_{\mathrm{S}}^{\perp}(E_{i})}\mathrm{e}^{-\mu_{\mathrm{air}}\rho_{\mathrm{air}}(d_{4}-d_{2}-d_{5})}.\label{eq:I_0L}
\end{eqnarray}
The count rates $I_{\mathrm{L}}$, $I_{\mathrm{S}}^{||}$, and $I_{\mathrm{S}}^{\perp}$
are extracted from the spectra in Figs. 3 and 4 as described above.
Table \ref{tab:Flux_results}  presents a summary of the determined fluxes of the Ga K$\alpha$
and In K$\alpha$ lines of the LMJ for two different power settings.
To be able to compare the flux with that of other sources we calculate
the brilliance for the LMJ for Ga K$\alpha_{1}$ and In K$\alpha_{1}$
with 200 W electron beam power. We obtain an estimated brilliance of $\sim 5\times10^{10}\,\mathrm{(s\, mrad^{2}\, mm^{2}\,0.1\%BW})^{-1}$
for Ga K$\alpha_{1}$ and $\sim 2\times10^{9}\,\mathrm{(s\, mrad^{2}\, mm^{2}\,0.1\%BW})^{-1}$ for In K$\alpha_{1}$, assuming a projected circular source of $20\,\mu\text{m}$ diameter, a K$\alpha_{1}$ to K$\alpha_{2}$ intensity ratio of 2:1, and by scaling the natural line width\cite{Krause1979} of 2.6 eV (Ga K$\alpha_{1}$) or 10.6 eV (In K$\alpha_{1}$).

Because target and detector specific parameters with potentially large
uncertainties cancel out, the uncertainties of the determined fluxes
are strongly reduced. 
Eq. (\ref{eq:I_0L}) allows for an analytic determination of the photon flux. Thus, the resulting uncertainties are derived according to the variance formula considering the propagation of uncertainties of the input parameters. An overview of the relative uncertainties of all used parameters is given in Table \ref{tab:uncertainties}. Largest contributions to the overall uncertainties are from the determination of the count rates \(I_{L} (E_i)\) at the LMJ due to background subtraction and the definition of a region of interest (Fig. \ref{fig:LMJ_glass_spec} (b) and (c)). Especially the overlapping peaks in the spectral region of In K\(\alpha\) lead to an estimated uncertainty of 8\% of the inelastic count rate. Measurement times were chosen in all experiments such that the relative statistical error is below 0.5\%. 

\begin{table}
 \caption{ \label{tab:Flux_results} Experimentally determined photon flux \(I_{0,LMJ}\) per solid angle of the LMJ for the characteristic K\(\alpha\)-lines of the anode material. The experimental value is determined using Eq. (\ref{eq:I_0L}). }
 \begin{ruledtabular}
 \begin{tabular}{l c c c c}
line   &	$E$ / keV            & power / W  & \(\frac{I_{0,L}^{U}}{\Omega_{L}}\) / s\(^{-1}\)sr\(^{-1} \times 10^{12}\)  \\ \hline
Ga K\(\alpha\)  &  9.25\footnote{according to ref.\cite{Deslattes2003}} &  200.0\footnote{nominal voltage 70 kV, current 2.857 mA}     &  \(6.0(5)\) \\
                &                           &  457.1\footnote{nominal voltage 160 kV, current 2.857 mA. We note that these settings are above the specifications of the manufacturer and may damage the source.}    &  \(12.4(9)\) \\ \hline
In K\(\alpha\)  & 24.21\(^{\text{a}}\) &  200.0\(^{\text{b}}\)     &  \(0.38(4)\) \\ 
                &                           &  457.1\(^{\text{c}}\)    &  \(1.77(17)\)
 \end{tabular}
 \end{ruledtabular}
 \end{table}

\begin{table}
\centering
 \caption{\label{tab:uncertainties} Uncertainty contributions to the overall uncertainty of the LMJ photon flux per solid angle for Ga K\(\alpha\) and In K\(\alpha\). }
  \begin{ruledtabular}
 \begin{tabular}{l c c l}
parameter           & \multicolumn{2}{l}{rel. uncertainty / \%}  & comment  \\
                    & Ga K\(\alpha\)& In K\(\alpha\)    &                    \\ \hline
\(I_{L}(E_i)\)&      3        & 8                 & detected count rate LMJ  \\
\(I^{\parallel, \perp}_{S}(E_i)\)    &      2        & 2                 & detected count rate BAMline \\
\(I^{P}_{D,S}(E_0)\)  & 1             & 1                 & determined flux BAMline \\
d\(_4\)-d\(_2\)     & 0.7             & 0.05                 & distance of air absorption \\
\(\rho_{\text{air}}\) & 0.4           & 0.03                 & density of air \\
\(\mu_{\text{air}}\)& 4            & 0.3                & attenuation coefficient of air \\
\(\Omega_{L}\)    & 5             & 5                 & solid angle of the LMJ \\\hline
\(I_{0,L}^U (E_0)/\Omega_L\) & 8             & 10                & determined flux per solid angle \\
 &            &                 & of the LMJ
 \end{tabular}
 \end{ruledtabular}
 \end{table}

In summary, we experimentally determined the photon flux per solid angle of two characteristic emission lines of a LMJ by referencing the unpolarized radiation of the LMJ to a well-known polarized synchrotron radiation beamline. 
High accuracy is achieved since knowledge of any type of X-ray interaction cross section is not needed as compared to other indirect methods. A calibrated SDD regarding its detection efficiency or the solid angle of detection is not needed since these quantities cancel given that the same setup is used at the synchrotron radiation source and the LMJ. The key prerequisites of our method are access to monochromatic X-rays in the desired energy range and a calibrated photo diode. Once a suitable scattering target has been measured and the scattering count rates have been related to the incoming photon flux, the compact setup allows for an easy transfer to numerous laboratory or medical X-ray sources. 







\section*{Acknowledgements}
The collaboration of PTB and HZB leading to this work was supported by the EMPIR project HyMet. The financial support of the EMPIR program is gratefully acknowledged. It is jointly funded by the EMPIR and participating countries within the European Association of National Metrology Institutes (EURAMET) and the European Union. We thank R. Häfner, PTB, for the construction of the experimental end-station and G. Wagener, HZB, for technical support in the LIMAX laboratory. 

\bibliography{bib}

\end{document}